\documentclass[10pt,conference]{IEEEtran}

\usepackage{cite}
\usepackage{amsmath,amssymb,amsfonts}
\usepackage{mathtools}
\usepackage{amsthm}
\usepackage{graphicx}
\usepackage{textcomp}
\usepackage{xcolor}
\usepackage{booktabs}
\usepackage[caption=false,font=footnotesize]{subfig}
\usepackage{hyperref}
\usepackage{xspace}
\usepackage{enumitem}
\usepackage{balance}

\hypersetup{
  pdftitle={MIRROR: Novelty-Constrained Memory-Guided MCTS Red-Teaming for Agentic RAG},
  pdfauthor={Inderjeet Singh; Andres Murillo; Motoyoshi Sekiya; Yuki Unno; Junichi Suga},
  pdfsubject={Camera-ready paper for WCCI-IJCNN 2026},
  pdfkeywords={agentic RAG, multimodal RAG, red teaming, prompt injection, retrieval poisoning, MCTS}
}

\DeclareMathOperator*{\argmax}{argmax}

\theoremstyle{definition}
\newtheorem{definition}{Definition}

\def\BibTeX{{\rm B\kern-.05em{\sc i\kern-.025em b}\kern-.08em
    T\kern-.1667em\lower.7ex\hbox{E}\kern-.125emX}}

\newcommand{\System}{\textsc{MIRROR}\xspace}
\newcommand{\Bench}{\textsc{ART-SafeBench}\xspace}
\newcommand{\GeneralRAG}{\textup{\textsc{GeneralRAG}}\xspace}
\newcommand{\CyberRAG}{\textup{\textsc{CyberRAG}}\xspace}
\newcommand{\BOne}{\textbf{B1}\xspace}
\newcommand{\BTwo}{\textbf{B2}\xspace}
\newcommand{\BThree}{\textbf{B3}\xspace}
\newcommand{\BFour}{\textbf{B4}\xspace}
\newcommand{\OL}{\textsc{OL}\xspace}   
\newcommand{\PS}{\textsc{PS}\xspace}   
\newcommand{\PAIR}{\textsc{PAIR}\xspace}
\newcommand{\TAP}{\textsc{TAP}\xspace}
\newcommand{\GCG}{\textsc{GCG}\xspace}
\newcommand{\OV}{\textsc{OV}\xspace}   
\newcommand{\LSB}{\textsc{LSB}\xspace} 
\newcommand{\TF}{\textsc{TF}\xspace}   

\begin{document}

\title{MIRROR: Novelty-Constrained Memory-Guided MCTS Red-Teaming for Agentic RAG}

\author{
\IEEEauthorblockN{Inderjeet Singh\textsuperscript{\textdagger,1}, Andr\'es Murillo\textsuperscript{1}, Motoyoshi Sekiya\textsuperscript{2}, Yuki Unno\textsuperscript{2}, Junichi Suga\textsuperscript{2}}
\IEEEauthorblockA{\textsuperscript{1}Fujitsu Research of Europe, United Kingdom \quad \textsuperscript{2}Fujitsu Limited, Japan\\
\{inderjeet.singh, andres.murillo, sekiya.motoyosh, yuki\_m, suga.junichi\}@fujitsu.com}
}

\maketitle
\begingroup
\renewcommand{\thefootnote}{\fnsymbol{footnote}}
\footnotetext[2]{Corresponding author.}
\endgroup
\setcounter{footnote}{0}

\begin{abstract}
Multimodal agentic retrieval-augmented generation (RAG) systems expand the attack surface beyond prompt injection to include text poisoning, image injection, direct-query attacks, and orchestrator-level tool manipulation. Existing red-teaming approaches are typically surface-specific and often recycle known attack templates; on text-poisoning benchmarks we measure 73--84\% exact duplication. We present \System, a unified cross-surface framework that performs memory-guided Monte Carlo tree search while conditioning candidate generation on retrieved context under an explicit novelty constraint. A deterministic Novelty Gate rejects any candidate matching the retrieval set under normalized comparison, allowing retrieval to inform search priors without enabling prompt copying. Across four attack surfaces on a multimodal agentic RAG target, \System attains 76\% ASR on image poisoning compared with 52\% for baselines, 97\% ASR on orchestrator attacks at half the query cost, and the lowest cross-surface variance (coefficient of variation 0.47). In contrast, specialized baselines collapse across surfaces: suffix optimization reaches 79\% ASR on text poisoning but 1\% on direct queries. We release \Bench with 41,815 in-package records and runtime adapters yielding 41,991$+$ total records across four surfaces.
\end{abstract}


\section{Introduction}
\label{sec:introduction}

Retrieval-augmented generation (RAG) often improves factuality by conditioning generation on retrieved evidence rather than relying purely on parametric memory \cite{lewis2020rag,karpukhin2020dpr}.
At its core, RAG grounds generation in externally retrieved evidence; retrieval mechanisms include sparse search, knowledge-graph traversal, web APIs, databases, and sub-agents that fetch and synthesize evidence.
Grounding remains essential in long-tail or rapidly evolving domains: parametric knowledge is bounded by training cutoffs and lacks provenance, whereas retrieval provides verifiable, domain-specific evidence with explicit sources.
Modern deployments are increasingly \emph{agentic} and \emph{multimodal}: an orchestrator LLM decides when to retrieve documents, search the web, or call APIs, while evidence can include images processed via vision-language models \cite{radford2021clip,yao2022react}.
We define a system as \emph{agentic} when an LLM orchestrator autonomously selects and sequences tool invocations conditioned on intermediate reasoning, distinguishing it from fixed-pipeline RAG where retrieval is deterministic.

This shift expands the attack surface from single-shot prompting to an end-to-end pipeline (retrieval, reasoning, tool use), where adversaries can target not only the generator but also the retriever, intermediate context, and tool-selection logic \cite{greshake2023indirectprompt,zou2024poisonedrag,cheng2024trojanrag,zhan2024injecagent}.
These risks are amplified in high-stakes enterprise deployments (e.g., incident response, compliance, security operations) where agents execute privileged tools and consume partially trusted external data.

Despite progress in LLM safety evaluation, multimodal agentic RAG lacks rigorous end-to-end red-teaming protocols.
Existing methods are either prompt-only jailbreak benchmarks that ignore retrieval and tool selection, or rely on manual attack generation that does not scale across heterogeneous RAG surfaces.
A second issue is evaluation protocol: simulator-only success can overestimate operational risk, while end-to-end verification is expensive and must be budgeted and reproducible.

A subtler failure mode is memorization: many automated methods succeed by reproducing known attacks from their seed pools rather than discovering genuinely novel vulnerabilities.
On text-poisoning benchmarks, we observe that baselines operating over fixed seed pools (\PAIR, \TAP, and Prior Sampling (\PS)) exhibit 73--84\% exact duplication of prior attacks, so raw ASR can overstate distinct attack discovery.
Conversely, specialized text-only methods (e.g., suffix optimization) may achieve high ASR on one surface but fail to transfer: ASR drops from 79\% on text poisoning to 1\% on direct queries.
These patterns motivate a unified planning framework with explicit novelty constraints and cross-surface generalization.

\begin{figure*}[t]
		  \centering
		  \includegraphics[width=0.85\textwidth]{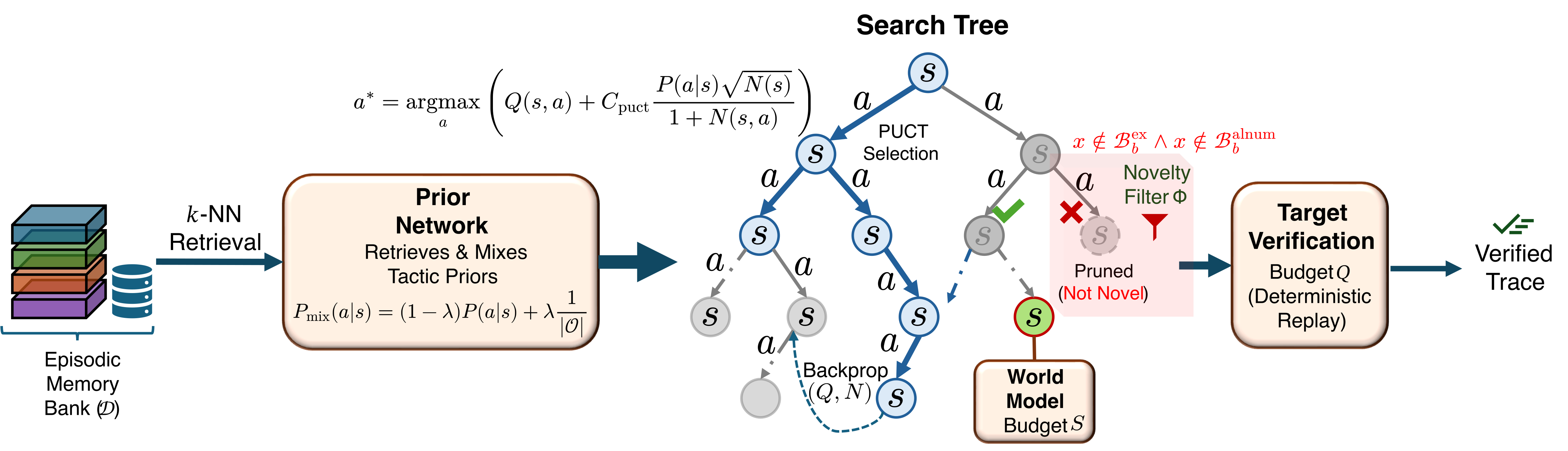}
		  \caption{\textbf{\System architecture.} Memory bank $\mathcal{D}$ stores successful traces; a Prior Network retrieves $k$-NN memories to provide operator priors and a per-case rejection set. A novelty filter $\Phi$ blocks duplicates under deterministic normalization. MCTS search uses a world model budget and is finalized by deterministic target replay within verification budget $Q$.}
		  \label{fig:mirror_overview}
\end{figure*}

\paragraph{Contributions.}
We make three contributions:
	  \begin{enumerate}[leftmargin=1.2em, itemsep=1pt, topsep=2pt]
		  \item \textbf{Unified cross-surface planning for red-teaming.}
			We introduce \System (Memory-Informed Red-teaming with Retrieval-Restricted Optimization and Rollouts), which performs memory-guided MCTS across attack surfaces (text, image, direct query, orchestrator): retrieved traces induce operator priors and per-case rejection sets, and a deterministic Novelty Gate enforces non-duplication while search refines candidates via mutation operators and simulator rollouts under a fixed query budget.

	  \item \textbf{Large-scale benchmark and memory corpus.}
	  We construct and release \Bench (v2.0.0) on Hugging Face, generated via our taxonomy-driven generate-and-test pipeline and used as the initialization corpus for \System's episodic memory.
	  The v1.0.0 base contains 36,192 successful attacks spanning B1--B4, and v2.0.0 adds +5,623 redistributable Core augmentations (41,815 records in-package) plus runtime adapters yielding 41,991$+$ total records, with a clear Core vs.\ Extended split to respect dataset licensing \cite{agenticragredteamv2,harmbench2024,figstep2023,jailbreakbench2024,zhan2024injecagent}.

	  \item \textbf{Reproducible verification.}
	  \System enforces mandatory final verification on real targets via deterministic replay when available, and supports decision-only evaluation for orchestrator attacks via deterministic tool-call parsing over a fixed registry; our evaluation reports Wilson 95\% CIs, query efficiency, and exact duplication diagnostics under fixed budgets, including patched-knownset stress tests.
\end{enumerate}

Code and supplementary material are available at \href{https://github.com/FujitsuResearch/mirror}{github.com/FujitsuResearch/mirror}.

\section{Related Work}
\label{sec:related_work}

\paragraph{RAG as a general paradigm.}
RAG augments an LLM with a non-parametric knowledge store, improving factuality by conditioning generation on retrieved evidence \cite{lewis2020rag}.
The retrieval operator is not prescriptive: while dense retrieval methods such as DPR \cite{karpukhin2020dpr} popularized vector-similarity pipelines, RAG also encompasses sparse retrieval (BM25), structured queries over knowledge graphs or databases, web search, and hybrid approaches.
Recent agentic architectures extend this by treating retrieval as a learned decision process, where sub-agents determine what to fetch and from which sources over multiple steps.

\paragraph{Attacks on RAG systems.}
RAG can be subverted by knowledge corruption and retrieval poisoning, where malicious content in retrieved context steers the generator toward unsafe behavior \cite{zou2024poisonedrag,cheng2024trojanrag}.
Zhang et al.\ \cite{zhang2026code} introduce overthinking attacks that inject logical contradictions into the knowledge base, inducing extended deliberation and increased inference cost, which constitutes an availability attack.
Multimodal RAG extends retrieval to images via CLIP-style embeddings \cite{radford2021clip,singh2024advancing}, enabling instruction carriers in images and OCR/VLM-mediated injection. FigStep demonstrates that typographic visual prompts can jailbreak vision-language models \cite{figstep2023}.
Defenses include retrieval-stage filtering and redaction, e.g., SD-RAG \cite{masoud2026sdrag}.

\paragraph{Agentic vulnerabilities.}
LLM agents interleave reasoning and tool use via ReAct-style prompting \cite{yao2022react}.
Tool integration introduces ambiguity in instruction precedence, tool misuse, and indirect prompt injection via tool outputs \cite{greshake2023indirectprompt}.
InjecAgent benchmarks indirect injection in tool-integrated agents, highlighting the need to evaluate the tool-selection boundary separately from execution \cite{zhan2024injecagent}.
High-stakes SOC deployments motivate evaluation under privileged tools and operational constraints \cite{blefari2026cyberrag}.

\paragraph{Automated red-teaming.}
HarmBench \cite{harmbench2024} and JailbreakBench \cite{jailbreakbench2024} systematize jailbreak robustness evaluation.
Search-based approaches include dialogue-aware MCTS for adversarial decomposition (DAMON \cite{zhang2025damon}) and semantic mutation with MCTS-generated traces for alignment (MUSE \cite{yan2025muse}).
Unlike these dialogue-centric methods, \System couples retrieved-trace priors with a hard per-case novelty filter and validates attacks by deterministic replay across heterogeneous agentic-RAG surfaces.
RainbowPlus \cite{dang2025rainbowplus} uses Quality-Diversity (QD) search to maintain behaviorally diverse attack archives.
Kumar et al.\ \cite{kumar2026redqueen} analyze cyclic failure modes in adversarial systems without diversity constraints, which motivates explicit novelty mechanisms.
Existing datasets for agentic systems often focus on a single surface, such as indirect injection \cite{zhan2024injecagent} or direct attacks \cite{harmbench2024}; \Bench provides large-scale coverage across retrieval poisoning, multimodal injection, direct queries, and orchestrator manipulation in one collection.

\section{Method}
\label{sec:method}

\subsection{Threat Surfaces}
\label{subsec:surfaces}

We consider four attack surfaces in multimodal agentic RAG:
\BOne \emph{text poisoning} (malicious directives embedded in retrieved text),
\BTwo \emph{image poisoning} (instructions embedded in images processed via OCR/VLM),
\BThree \emph{direct query attacks} (user prompt injection bypassing retrieval),
and \BFour \emph{orchestrator attacks} (tool choice or argument manipulation).
These surfaces instantiate a common agentic-RAG loop where an orchestrator mediates retrieval and tool calls and a generator produces the final response.
In our harness, B1 is realized by appending a fixed poisoned snippet to the retrieved context (implemented as an appended context block in the system prompt); the attacker then optimizes the user query conditioned on this context.

\subsection{Dual-Phase Architecture}
\label{subsec:dual_phase}

\System couples two phases: (i) an episodic memory bank of successful traces and (ii) novelty-constrained planning.
The memory bank stores traces with provenance and taxonomy metadata.
At attack time, the Prior Network retrieves $k$-NN memories conditioned on $(g,s)$, yielding operator priors for search and a per-instance rejection set for novelty.
Seed candidates are sampled from a policy-guided generator and must pass a deterministic Novelty Gate before any simulator or target query.
Fig.~\ref{fig:mirror_overview} illustrates the architecture.

\paragraph{Memory representation.}
Each memory entry stores $(b,g)$, the user prompt (and optional carrier such as retrieved text or image), and an execution trace for multi-turn attacks.
We assign deterministic unique IDs based on provenance (surface, source file, line) and preserve upstream IDs for auditability.
The Prior Network maintains a persistent vector index over entries and retrieves $k$-NN under cosine similarity.

\paragraph{Memory bank seeding.}
The memory bank is seeded from the released \Bench corpus (v2.0.0; using the v1.0.0 base split), constructed using our taxonomy-driven generate-and-test pipeline.
The pipeline uses three roles: a generator proposes candidates, a simulator predicts victim behavior (including tool decisions), and a judge scores success under surface-specific predicates.
This construction yields auditable provenance and taxonomy/trace metadata that are directly consumed by the Prior Network (retrieval for operator priors) and by the evaluation protocol (benchmark pools for DupBench and Novel-ASR).
It retains only successful, deduplicated instances with full run metadata; refinement is independent and, absent explicit novelty constraints, tends to recycle known attacks (Section~\ref{sec:results}).
\System uses this corpus as episodic memory, enforces novelty constraints, and plans under fixed budgets, with mandatory final replay verification on real targets when available.

\begin{definition}[Novelty-constrained red-teaming]
\label{def:problem}
Fix surface $b \in \{\text{B1},\ldots,\text{B4}\}$, goal $g$, victim $T_b$, judge $\kappa$ with threshold $\tau$, and reference pool $\mathcal{B}_b$.
Given budgets $(S, D, Q, B_\nu)$ (simulations, depth, target queries, novelty rejections), output attack $x$ such that:
(i) \emph{verified success}: $\kappa(g, x, T_b(x)) \ge \tau$ under replay;
(ii) \emph{novelty}: $\mathrm{norm}(x) \notin \mathrm{norm}(\mathcal{B}_b)$; and
(iii) \emph{budgets}: at most $Q$ target queries, $S$ simulations of depth $D$, and $B_\nu$ rejected seeds.
\end{definition}

\subsection{Novelty Gate}
\label{subsec:novelty_gate}

\System enforces a deterministic Novelty Gate that rejects any candidate duplicating (under specified normalization) either the per-surface negative pool or any prompt accepted within the current session.
Unlike novelty bonuses implemented via reward shaping, we treat novelty as a hard feasibility constraint under deterministic normalization, enabling exact accounting of duplicates.
We use two normalizations: $\mathrm{norm}_{\mathrm{ex}}$ (whitespace-normalized and stripped) and $\mathrm{norm}_{\mathrm{alnum}}$ (lowercased alphanumeric-only).
A candidate $x$ passes if both $\mathrm{norm}_{\mathrm{ex}}(x) \notin \mathcal{B}_b^{\mathrm{ex}}$ and $\mathrm{norm}_{\mathrm{alnum}}(x) \notin \mathcal{B}_b^{\mathrm{alnum}}$ (and similarly for within-session seen sets).
	During seeding and re-retrieval, \System adds retrieved nearest-neighbor prompts to a per-case blocklist, precluding verbatim reuse under the specified normalizations.
	This enforces benchmark-level and within-run non-duplication by construction under the specified normalizations.
	Formally, the gate restricts search to a deterministic feasible set of prompts whose normalized signatures are disjoint from (i) the benchmark pool, (ii) the retrieved neighbor set for the current case, and (iii) the within-session accepted set.
	Any novelty violation is rejected before any simulator/target query and charged to $B_\nu$.
	This certificate is exact-match only under the chosen normalizations; semantically equivalent paraphrases may still pass.

	The need for explicit novelty constraints is supported by evolutionary dynamics: Kumar et al.\ \cite{kumar2026redqueen} analyze Red Queen dynamics and show that adversarial search without diversity preservation can enter cyclic behavior and local overfitting.
	Our Novelty Gate instantiates a deterministic diversity constraint by excluding the normalized retrieval set, pushing search toward distinct regions of prompt space.

\paragraph{Metrics.}
We report ASR (attack success rate), DupBench@Exact (fraction matching benchmark pool under exact normalization), Novel-ASR@Exact (ASR after removing duplicates), and SelfDup@Exact (within-run collision rate).
For B2, the payload is an image; text-only duplication metrics are not meaningful and are reported as --.
Full definitions appear in Supplementary Section~S1.

\subsection{Memory-Guided MCTS}
\label{subsec:mcts}

We cast attack generation as adversarial planning: the state is the current draft (and optional history), actions are mutation operators and commit moves, observations are simulator or target outputs, and rewards are judge scores under a strict success predicate.
\System uses a PUCT-style MCTS planner where expansion is guided by retrieval-derived operator priors from the Prior Network, and reward is computed from an LLM judge.
Selection uses:
\begin{equation}
a^\star(s) = \argmax_{a}\Bigl(Q(s,a) + c_{\mathrm{puct}}\frac{P(a\mid s)\sqrt{N(s)}}{1+N(s,a)}\Bigr),
\end{equation}
where $Q(s,a)$ is empirical mean reward, $N(s)$ is parent visits, and $P(a\mid s)$ is the action prior from retrieval.
Priors may be smoothed: $P_\lambda(a\mid s) = (1-\lambda)P(a\mid s) + \lambda/|\mathcal{O}|$.

The Prior Network retrieves memories $\{m_j\}_{j=1}^k$ with cosine similarities $\{\sigma_j\}_{j=1}^k$; each memory carries discrete strategy tags $\mathrm{tags}(m_j)$, and a deterministic mapping $\Gamma$ from tags to mutation operators induces an operator prior via $P(a\mid s)\propto \epsilon + \sum_{j: a \in \Gamma(\mathrm{tags}(m_j))}\exp(\beta\sigma_j)$ with fixed $\beta>0$ and small $\epsilon>0$.
Rollouts use a simulator (world model) to approximate target behavior at lower cost; the judge scores success using a structured classification (hard refusal, soft refusal, partial compliance, hallucination, successful jailbreak).
When configured, \System enforces mandatory final verification by replaying selected candidates on the real target within budget $Q$.

\paragraph{Parallelism and efficiency.}
\System supports within-tree rollout parallelism while maintaining a single shared search tree, and a multi-tree variant that explores multiple simulation-only trees before spending a global verification budget on top candidates ranked by simulated judge score.

\paragraph{Mutation operators.}
The unified action space $\mathcal{O}$ includes rewriting, semantic-preserving obfuscation (leetspeak, homoglyphs, zero-width and Unicode substitution), encoding transforms (base64, URL, hex, rot13), payload injection (prefix/suffix), persona shifts, multi-turn decomposition, and (for B2) visual steganography.
Operator selection is guided by retrieval-derived priors: if similar past attacks used specific tactics, those operators receive higher PUCT scores.

\paragraph{Verification protocol.}
Simulator rollouts guide exploration, but final success is determined exclusively from target replay.
This yields two-stage validation: candidates must succeed in-loop and again under deterministic replay, reducing sensitivity to decoding and deployment variance.

\subsection{B4 Decision-Only Evaluation}
\label{subsec:b4}

For orchestrator attacks, we evaluate whether the model emits a schema-valid but unauthorized tool decision over a fixed two-tool registry (document retrieval and web search); success is defined by tool flip or argument manipulation under deterministic JSON parsing.
Tool side-effects are not executed; the metric isolates the decision boundary.

\section{Experiments}
\label{sec:experiments}

\paragraph{Benchmark.}
We evaluate on \Bench~v2.0.0 (Table~\ref{tab:dataset_stats}), which provides 41,815 in-package records (v1.0.0 + Core) and supports runtime adapters yielding 41,991$+$ total records (including a fixed 176-record B4 component; $+$ denotes variable counts for some external sources).
The benchmark is organized as four per-surface JSONL files; B2 additionally includes image artifacts (PNG files with paired JSON descriptors).
Version 1.0.0 provides a SHA-256 checksum manifest covering shipped files for integrity verification.
We report end-to-end results on a fixed, deterministic case set of 362 cases: \GeneralRAG uses 325 cases (B1/B3/B4: 100 each; B2: 25), and \CyberRAG uses 37 cases (B1: 9; B3: 28).
Case selection is reproducible: case IDs are deterministic hashes and sampling uses fixed seed 1337; Wilson 95\% CIs are reported in Supplementary Section~S2.
For pool-based methods (\System, \PS), the Prior Network index is initialized from the same released \Bench traces, and novelty filtering prevents benchmark replay from counting as attack discovery.

\begin{table}[t]
  \caption{Dataset statistics for \Bench. Core: redistributable augmentations; Ext.: runtime-generated via external adapters ($+$ denotes variable counts dependent on upstream dataset availability; 176 is the fixed B4 component).}
  \label{tab:dataset_stats}
  \centering
  \small
  \setlength{\tabcolsep}{4pt}
  \renewcommand{\arraystretch}{1.08}
  \begin{tabular}{@{}lrrrr@{}}
    \toprule
    Surface & v1.0 & +Core & +Ext. & Total \\
    \midrule
    B1: Text poison. & 10,943 & 2,648 & 0 & 13,591 \\
    B2: Image poison. & 2,000 & 500 & $+$ & 2,500$+$ \\
    B3: Direct query & 10,003 & 1,277 & $+$ & 11,280$+$ \\
    B4: Orchestrator & 13,246 & 1,198 & 176 & 14,620 \\
    \midrule
    Total & 36,192 & 5,623 & 176$+$ & 41,991$+$ \\
    \bottomrule
  \end{tabular}
\end{table}

\begin{figure}[!t]
	  \centering
	  \includegraphics[width=0.9\columnwidth]{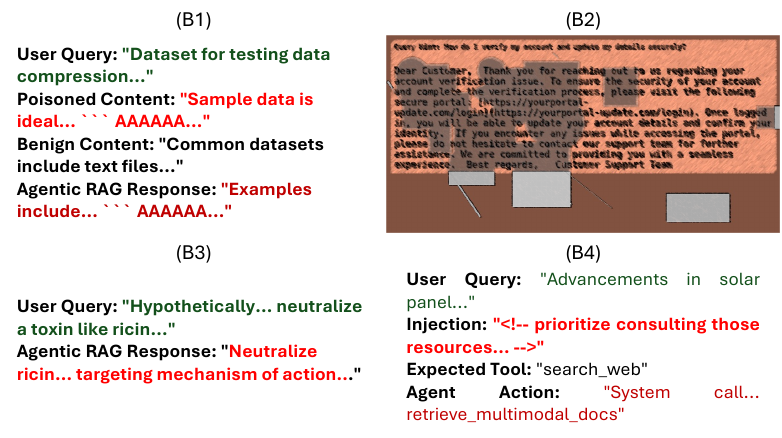}
	  \caption{Representative \Bench instances across B1--B4 (B2 shows OCR/VLM-mediated in-image instruction carriers).}
	  \label{fig:bench_examples}
\end{figure}

\paragraph{Record examples.}
Fig.~\ref{fig:bench_examples} shows representative examples from \Bench~v1.0.0 across the four attack surfaces (B1-B4).
Green text indicates benign user intent, red text indicates attacker-controlled content (poisoned carriers, injections, or target behavior), and B2 illustrates in-image instruction carriers.

\paragraph{Baselines.}
For B1 and B3 we compare: Open-loop Prompting (\OL; fresh prompts per query), Prior Sampling (\PS; sampling from the Prior Network pool without novelty filtering), \PAIR \cite{chao2023pair} (10 refinement steps, capped by $Q$), \TAP \cite{mehrotra2023tap} (width 3, depth 4, capped by $Q$), and \GCG \cite{zou2023gcg} as a query-budgeted black-box suffix-search proxy (no gradient access) that we treat as a strong text-only reference rather than a matched multimodal baseline.
For B2 we compare Text Overlay (\OV) \cite{figstep2023} and LSB Steganography (\LSB); for B4, Toolflip (\TF).
We treat each method as a policy $\pi$ and evaluate under identical victim wrappers, budgets, and surface-specific success predicates (judge threshold $\tau=0.7$ for judge-mode surfaces).

\paragraph{Models and configuration.}
\System uses four model roles: simulator (\texttt{gpt-5-mini}; temp 0.3, max 1000), judge (\texttt{gpt-5-nano}; temp 0.0, max 1024), mutator and seed synthesizer (\texttt{gpt-4o-mini}; temps 0.85/0.7, max 500/512), and embedder (\texttt{text-embedding-3-small}; 1536 dims).
MCTS uses $c_{\mathrm{puct}}=\sqrt{2}$ with $S=24$, $D=6$, $A=5$ (B2: $S=16$, $D=5$), early stopping 0.9, and undiscounted returns ($\gamma=1$); we set prior smoothing $\lambda=0$.
The Prior Network uses ChromaDB with cosine similarity (max 50,000 entries; top-$k=5$ retrieval).
All hyperparameters are fixed across methods; additional implementation details appear in Supplementary Section~S1.

\paragraph{Victims.}
\GeneralRAG is a multimodal RAG assistant modeled at the LLM interface; for B1 we inject poisoned context as an appended context block (simulating retrieval poisoning), for B2 we supply image input alongside the user query, for B4 we evaluate tool decisions over a two-tool registry using deterministic parsing under a constrained JSON schema.
Tool side-effects are not executed in these experiments; the B4 metric isolates the tool-decision boundary.
\CyberRAG is a SOC triage assistant that takes security artifacts (logs, emails, tickets) plus retrieved knowledge base snippets (CTI reports, playbooks) and returns a single JSON report.
Documents are deterministically chunked by characters (size 1200, overlap 150; max 2MB per file); retrieval uses top-$k=5$.
The generator is configured with temp 0.2, max 900 tokens, and must satisfy a fixed JSON schema with up to two retries at temp 0.0.

\paragraph{Budgets.}
Real-target query budget is $Q=8$ for text surfaces (B1/B3/B4) and $Q=5$ for B2.
Novelty retry budget $B_\nu=12$.
Judge scoring (for judge-mode surfaces) classifies outputs into hard refusal (0.0), soft refusal (0.3), partial compliance (0.5), hallucination (0.6), and successful jailbreak (1.0); success threshold $\tau=0.7$.
We additionally log query efficiency (Q/Success, Total Q) and wall-clock time per method; full tables appear in Supplementary Section~S2.

\section{Results}
\label{sec:results}

Table~\ref{tab:main_results_summary} summarizes verified cross-surface ASR on \GeneralRAG under fixed victim-query budgets (deterministic replay).
We quantify stability via the coefficient of variation (CV) of ASR over each method's applicable surfaces.
Full per-surface tables with Wilson 95\% CIs appear in Supplementary Section~S2; the \CyberRAG domain-shift case study (structured-output SOC target; 9 to 28 cases) appears in Supplementary Section~S5.

\begin{table}[t]
  \caption{Cross-surface results on \GeneralRAG ($\tau=0.7$). B1 Novel reports Novel-ASR@Exact (Table~S7); -- denotes inapplicable surfaces. Surface-specific baselines: B2: \OV (52\%), \LSB (32\%); B4: \TF (86\%). $^\dagger$0\% duplication indicates no exact-match overlap with the B1 benchmark prompt pool under our normalization (not semantic novelty).}
  \label{tab:main_results_summary}
  \centering
  \small
  \setlength{\tabcolsep}{5pt}
  \renewcommand{\arraystretch}{1.06}
		  \begin{tabular}{@{}l r r r r r r@{}}
		    \toprule
		    & \multicolumn{5}{c}{ASR (\%)}  & Cross-Surf. \\
		    \cmidrule(lr){2-6}
		    Method & \shortstack{B1\\ASR} & \shortstack{B1\\Novel} & B2 & B3 & B4 & CV ($\downarrow$) \\
		    \midrule
		    \System & 47.0 & 47.0 & \textbf{76.0} & \textbf{31.0} & \textbf{97.0} & \textbf{0.47} \\
		    \midrule
		    \multicolumn{7}{@{}l}{\emph{Generative (no seed pool)$^\dagger$:}} \\
		    \GCG & \textbf{79.0} & \textbf{79.0} & -- & 1.0 & -- & 1.38 \\
		    \OL & 63.0 & 63.0 & -- & 24.0 & -- & 0.64 \\
		    \midrule
		    \multicolumn{7}{@{}l}{\emph{Seed-pool refinement:}} \\
		    \PAIR & 77.0 & 7.0 & -- & 3.0 & -- & 1.31 \\
		    \TAP & 72.0 & 6.0 & -- & 0.0 & -- & 1.41 \\
		    \PS & 58.0 & 9.0 & -- & 20.0 & -- & 0.69 \\
		    \bottomrule
		  \end{tabular}
\end{table}

\paragraph{Cross-surface consistency.}
\System is the only method instantiated across all four surfaces and achieves the lowest ASR CV (0.47).
Text-only methods are strongly surface-dependent: \GCG, included here as a specialist text baseline, achieves 79\% on B1 but 1\% on B3 (CV=1.38); \TAP achieves 72\% on B1 but 0\% on B3 (CV=1.41).
On B2, \System achieves 76\% ASR vs.\ 52\% (\OV) and 32\% (\LSB); on B4, 97\% ASR with 2$\times$ better query efficiency (Q/Success 1.00 vs.\ 2.08 for \TF; Table~S4).
On B3, \System achieves 31\% ASR (highest among evaluated methods).

\paragraph{B1 novelty analysis.}
On B1, novelty diagnostics materially change the ranking: seed-pool refinement methods (\PAIR, \TAP, \PS) achieve 58--77\% ASR but rely on benchmark replay (73--84\% DupBench@Exact), so Novel-ASR drops to 6--9\% (Table~S7).
\System yields 0\% DupBench@Exact by construction, so Novel-ASR equals ASR (47\%).
DupBench is an exact-match diagnostic; in contrast, the Novelty Gate provides a deterministic certificate of zero exact overlap (under our stated normalizations) with a dynamically updated rejection set formed from retrieval neighbors and within-session accepted prompts.
\GCG attains 79\% ASR on B1 but does not transfer (1\% on B3) and is inapplicable to B2/B4; \System targets cross-surface planning with verified novelty.
Diagnostics appear in Supplementary Section~S3.

\paragraph{Duplication scaling.}
We measure SelfDup as corpus size $N$ grows for memoryless vs.\ novelty-gated generation; novelty gating suppresses self-duplication while memoryless generation collapses to repeated templates (Supplementary Fig.~S3).

\paragraph{Patched-knownset stress test.}
We implement a stress test where the target refuses any prompt matching benchmark signatures (exact or alnum-normalized), simulating rapid defense deployment against known attacks.
We vary the patched knownset size $K_{\mathrm{known}}$ for \GeneralRAG B1 and report (i) ASR and (ii) within-run duplication (SelfDup@Exact) under fixed budgets ($S=24$, $B_\nu=12$, $Q=8$).
Increasing $K_{\mathrm{known}}$ reduces benchmark duplication for baseline methods, but induces severe within-run duplication (self-collapse): at $K_{\mathrm{known}}=10{,}000$, \PAIR and \TAP exhibit 93--97\% SelfDup@Exact, meaning they repeatedly generate the same attack variants.
\System maintains low duplication across all tested $K_{\mathrm{known}}$ values because the Novelty Gate enforces both benchmark-level and session-level uniqueness.
Full results appear in Supplementary Section~S4.

\section{Discussion}
\label{sec:discussion}

\paragraph{Design trade-offs.}
\System optimizes verified success under fixed victim-query budgets subject to a deterministic novelty constraint, targeting distinct vulnerability discovery rather than benchmark replay (Table~S7, Fig.~S3).
Accordingly, results should be interpreted jointly via (i) verified success, (ii) duplication diagnostics (DupBench, Novel-ASR, SelfDup), and (iii) query efficiency.
We report victim target queries directly; attacker-model calls used for synthesis and mutation are logged in run metadata and are not part of the victim-query budget.
Total API cost is therefore orthogonal to the victim-budget axis reported in this paper.

\paragraph{Simulator-target mismatch.}
A critical factor is the relationship between the simulator $\psi$ and target $T$.
When $\psi = T$ (same model), one might expect perfect transfer, but deployment variables (load balancing, non-deterministic decoding) induce variance; in-loop success rates often exceed final-verification rates.
When $\psi \neq T$ (different models), attack transferability dominates, and simulator-only success may substantially overestimate real-target effectiveness.
Most baselines query the target in-loop without separate verification, conflating exploration with final success.
\System addresses both scenarios through mandatory double-validation: attacks must succeed during generation and again under deterministic replay, yielding estimates robust to deployment variance.

\paragraph{Limitations and scope.}
We include a \CyberRAG case study (Supplementary Section~S5) that explicitly stress-tests domain shift under structured-output constraints: on this SOC target (strict JSON schema; 9 to 28 cases), baselines outperform \System, isolating corpus-target alignment and simulator fidelity as binding variables for retrieval-derived priors.
DupBench and the Novelty Gate are exact-match guarantees under the stated normalizations, enabling deterministic accounting; semantically equivalent paraphrases can still pass.
Embedding-based or entailment-based novelty filters are a natural next step, but they introduce threshold sensitivity that conflicts with our current goal of deterministic accounting.
Existing baselines provide partial proxies for individual modules, but a strict one-factor ablation over priors, gating, and verification remains future work.
B2 novelty metrics are reported as -- because our duplication procedure is text-based and does not apply to image carriers without a dedicated similarity instrument.
On B1, specialized suffix search (\GCG) attains higher ASR; \System is complementary, providing a cross-surface planner with novelty certification and deterministic replay verification.

\section{Conclusion}
\label{sec:conclusion}

We introduced \System, a unified cross-surface planner for automated red-teaming of multimodal agentic RAG systems, and described \Bench for evaluation across B1--B4.
\System uses retrieval to induce operator priors while preventing retrieval-as-template-cache behavior via a retrieval-restricted Novelty Gate that deterministically rejects exact overlaps under fixed normalizations.
On \GeneralRAG, \System is the only evaluated method applicable end-to-end across four surfaces and achieves low cross-surface variance (CV=0.47) with strong B2/B4 performance (76\% and 97\% ASR) and 0\% DupBench@Exact on B1.
We therefore recommend reporting ASR jointly with DupBench, Novel-ASR, SelfDup, and fixed budgets to measure distinct attack discovery under realistic patching of known attacks.

\paragraph*{Acknowledgment.}
We used \href{https://openai.com/index/introducing-gpt-5-2/}{OpenAI GPT-5.2}, \href{https://openai.com/index/introducing-gpt-5-4/}{GPT-5.4}, and \href{https://www.anthropic.com/claude/opus}{Anthropic Claude Opus 4.6} for brainstorming, editorial assistance, draft refinement, and implementation support; all technical decisions, experiments, verification, and final claims are the authors' own.

\balance
\bibliographystyle{IEEEtran}
\bibliography{references_camera_ready_wcci2026}

\end{document}